\documentclass[oneside,brazil,english,aps,preprint]{elsart}
\usepackage[T1]{fontenc}
\usepackage[latin9]{inputenc}
\usepackage{float}
\usepackage{textcomp}
\usepackage{relsize}
\usepackage{amsmath}
\usepackage{graphicx}
\usepackage{amssymb}

\makeatletter

\DeclareRobustCommand{\lyxmathsym}[1]{\ifmmode\begingroup\def\b@ld{bold}
  \def\rmorbf##1{\ifx\math@version\b@ld\textbf{##1}\else\textrm{##1}\fi}
  \mathchoice{\hbox{\rmorbf{#1}}}{\hbox{\rmorbf{#1}}}
  {\hbox{\smaller[2]\rmorbf{#1}}}{\hbox{\smaller[3]\rmorbf{#1}}}
  \endgroup\else#1\fi}

\providecommand{\tabularnewline}{\\}

\makeatother

\usepackage{babel}

\begin{document}
\begin{frontmatter}

\title{Screening masses in quenched $\left(2+1\right)d$ Yang-Mills theory:
universality from dynamics?}

\author{Rafael B. Frigori}

\ead{frigori@utfpr.edu.br}

\address{Universidade Tecnológica Federal do Paraná, \\
Rua XV de Novembro 2191, 85902-040 Toledo (PR), Brazil}
\begin{abstract}
We compute the spectrum of gluonic screening-masses in the $0^{++}$
channel of quenched 3d Yang-Mills theory near the phase-transition.
Our finite-temperature lattice simulations are performed at scaling
region, using state-of-art techniques for thermalization and spectroscopy,
which allows for thorough data extrapolations to thermodynamic limit.
Ratios among mass-excitations with the same quantum numbers on the
gauge theory, 2d Ising and $\lambda\phi^{4}$ models are compared,
resulting in a nice agreement with predictions from universality.
In addition, a gauge-to-scalar mapping, previously employed to fit
QCD Green's functions at deep IR, is verified to dynamically describe
these universal spectroscopic patterns. \end{abstract}
\end{frontmatter}

\section{Introduction}

The lattice formulation of gauge theories \cite{LatticeQCD} as QCD
allows for \textit{ab initio} systematically well-controlled numerical
investigations of most theoretical key aspects, like confinement and
chiral symmetry breaking \cite{Alkofer_Confinement}, at non-perturbative
regime. For instance, at high temperatures or densities when deconfinement
phase-transition takes place in non-abelian gauge theories, Monte
Carlo simulations have been essential to unveil how matter behaves
under extreme thermodynamical conditions \cite{Karsch_QGP_LQCD}. 

Hence, quantum field theories (QFT) on the lattice can be understood
as classical models of statistical mechanics. Some enlightening interconnections
usually appear from finite temperature studies \cite{Stat_Field_Theory},
as in the case of a long-standing conjecture --- designed upon general
symmetry arguments --- from Svetistky and Jaffe \cite{Svet&Yaffe}
relating universal properties from finite-temperature $SU(N)$ gauge-theories
in $\left(d+1\right)$ dimensions to $d-$dimensional $Z_{N}-$spin
systems at criticality. This has been corroborated by computations
of critical (static and dynamic) exponents, universal amplitudes,
and correlation functions: see for example, respectively \cite{Critical_exponents}\cite{Universal_amplitude_ratios}\cite{Correlations_at_Criticality}.

Still, under the spontaneous breaking of global $Z_{N}$ center-symmetry,
$SU(N)$ gauge theories were shown --- in 4-dimensions and $N\leq3$
--- to exhibit a rich spectrum of gluonic screening-masses, whose
ratios among equivalent excitations are shared by $Z_{N}-$spin systems
\cite{Fiore,Falcone}. Despite the effectiveness of universality-based
reasoning, investigations on symplectic gauge theories compel for
a still elusive dynamical description of deconfinement \cite{GeneralGroups},
instead of relying completely on $Z_{N}-$invariance as a physical
fingerprint. 

New perspectives have appeared over the last years, such as a unified
picture of confinement emerging from various analysis of Green\textasciiacute{}s
functions of non-abelian gauge theories \cite{Alkofer_Confinement,AguilarNatale,AlkoferHubberSchwenzer_decoupling,QCD_Props_Constraints_ACTM,Mapping_Theorem_Frasca}.
While analytical (decoupling) solutions of Dyson-Schwinger equations
\cite{AguilarNatale,AlkoferHubberSchwenzer_decoupling} and gauge-to-scalar
mappings \cite{Mapping_Theorem_Frasca} predict non-vanishing gluon
propagators at deep infrared for $d>2$, lattice computations provide
evidences of gluons having dynamically-generated masses \cite{QCD_Props_Constraints_ACTM}. 

These features stay unchanged at finite-temperatures, which supports
a strongly-coupled plasma at the vicinity of phase-transition, thus
agreeing with dimensional-reduced theories \cite{Cucchieri_Screening}
and (universal) results from AdS/QCD correspondence \cite{AdS_QCD}.
Additionally, properties derived from gauge-dependent gluon propagators
would also be compared with spectroscopical analysis employing gauge-invariant
correlators \cite{Maezawa_Screnning_Masses} (free from Gribov ambiguities
\cite{Gribov_copies_Cucchieri}), which may deepen physical insights
into QFT- thermodynamics. For a more comprehensive review on gluonic
bound-states using methods from spectroscopy, at zero and finite-temperatures,
see also \cite{Review_Glueball_Spectroscopy}.

In this article, we investigate through Monte Carlo simulations the
gluonic screening-spectrum of quenched three-dimensional Yang-Mills
(YM) theory on the lattice. State-of-art techniques for spectroscopy,
such as the variational-method \cite{Variational_Method} and recursively-smeared
interpolators \cite{Fiore,Falcone}, are employed to compute excitations
on the $0^{++}$ channel. Using an improved thermalization algorithm
\cite{MHB} we carry out simulations at asymptotic region with ameliorated
critical-slowing-down (CSD) \cite{Sokal_Wolff}, UV-discretization
and IR-cutoff effects \cite{Strong_coupling_limit_QCD}. We compare
mass-ratios of YM theory and exactly-solvable bidimensional systems
in the same universality-class (i.e. the Ising and $\lambda\phi^{4}$
models) \cite{Svet&Yaffe}, therefore enabling us to check predictions
from universality \cite{Svet&Yaffe,Fiore,Falcone} and other dynamical
mechanisms \cite{GeneralGroups,Mapping_Theorem_Frasca}.

This work is organized as follows: in section II there is a review
of the major aspects of universality and its implications for thermal
excitations on gauge theories and spin-systems at the vicinity of
critical points. Section III outlines the general simulation setup,
as well as the modified heat-bath algorithm used for ensuring more
efficient thermalization of gauge fields. Spectroscopical techniques
for extracting screening masses are the topic of Section IV. Section
V presents numerical results and comparative data analysis. Conclusions
of the present work and its outlook are discussed in Section VI.

\section{Universality: field theories and spin-systems}

The finite-temperature formalism for $d-$dimensional gauge theories
on the lattice \cite{LatticeQCD,Stat_Field_Theory} uses an asymmetric
euclidean spacetime, whose physical volume $V=a^{d}\cdot L_{S}^{d-1}\cdot L_{T}$
has $L_{S}$ and $L_{T}$ as spatial and temporal (adimensional) lattice
sides, respectively. Thus, under the assumption $a\ll\left(a\cdot L_{T}\right)<\left(a\cdot L_{S}\right)$,
where $a$ is the (dimensionful) lattice spacing, one can set the
temperature of physical equilibrium as $T{}^{-1}=a\cdot L_{T}$. As
a consequence of the formalism, in the vicinity of phase transitions,
arguments from universality \cite{Svet&Yaffe} state that finite-temperature
QFTs for $SU(N)$ gauge groups in $\left(d+1\right)$ dimensions belong
to the same universality class of globally $Z_{N}$-invariant spin-systems
in $d$ dimensions. 

For the particular case of $SU(N=2)$ gauge theory, a second-order
(i.e. critical) deconfinement phase-transition is expected. This theory
can be implemented on the lattice by the Wilson action\begin{equation}
S_{W}\left[U\right]\equiv\beta\sum_{x}\sum_{\mu,\nu=1}^{d}\left\{ 1-\frac{1}{2}Re(TrP_{\mu\nu})\right\} ,\label{Wilson_Action}\end{equation}
using $SU(2)$ gauge-links $U_{\mu}\left(x\right)$ and the plaquette\begin{equation}
P_{\mu\nu}\equiv U_{\mu}\left(x\right)U_{\nu}\left(x+\hat{\mu}a\right)U_{\mu}^{-1}\left(x+\hat{\nu}a\right)U_{\nu}^{-1}\left(x\right),\label{Plaquete}\end{equation}
with lattice-coupling $\beta=2N/g_{s}^{2}a^{4-d}$ and gauge-field
coupling constant $g_{s}^{2}$. The action Eq.(\ref{Wilson_Action})
is invariant under transformations generated by center-elements of
the gauge-group, hence, spatial loops --- in contradistinction to
temporal ones --- are insensitive to temperature-induced symmetry
breaking. Then, a natural order parameter is given by a temporal loop,
the so called Polyakov loop $\bar{L}\equiv\left\langle L\left(x,y,z\right)\right\rangle =\left\langle Tr\prod_{n=1}^{n=L_{T}}U_{T}\left(x,y,z,an\right)\right\rangle $
\cite{Polyakov}, which measures the potential energy necessary to
free a quark \cite{Karsch_QGP_LQCD}.

On the other hand, the $3d$ Ising model is a well-known spin-system
that undergoes a second-order phase-transition, for $\beta_{I}=\beta_{critical}$,
due to a global $Z_{2}-$symmetry breaking universally related to
the finite-temperature $(3+1)d$ YM theory. It is described by the
action\begin{equation}
S_{I}=-\beta_{I}\sum_{<n,m>}s_{n}s_{m}.\label{Ising_Action}\end{equation}
In this same universality class is the $3d$ lattice-regularized $\phi^{4}$
theory, whose action is

\begin{equation}
S_{\phi}=-\beta_{\phi}\sum_{<n,m>}\phi{}_{n}\phi_{m}+\sum_{n}\phi{}_{n}^{2}+\lambda\sum_{n}\left(\phi{}_{n}^{2}-1\right)^{2}.\label{Phi4_Action}\end{equation}
In both cases, the order parameters are expectation values of fundamental
fields, e.g. the magnetization $(\bar{m)}$ or the v.e.v. $\left(\bar{\phi}\right)$
of the scalar-field, which behaves as the Polyakov loop $\left(\bar{L}\right)$
in the YM theory \cite{Fiore}.

The analogy among order-parameters $\bar{m},\bar{\phi}$ and $\bar{L}$
can also be extended to their correlation functions. The connected
correlation function $G_{C}$ is usually expressed as \begin{equation}
\begin{array}{c}
G\left(\left|r_{2}-r_{1}\right|\right)_{C}\equiv<O\left(r_{2}\right)O^{\dagger}\left(r_{1}\right)>-<O\left(r_{2}\right)><O^{\dagger}\left(r_{1}\right)>=\\
\underset{i}{\sum}A_{i}\left(e^{-m_{i}\left|r_{2}-r_{1}\right|}+e^{-m_{i}\left(L_{S}-\left|r_{2}-r_{1}\right|\right)}\right),\end{array}\label{Corr-Matrix}\end{equation}
which is a sum over each mass $\left(m_{i}\right)$ of the spectrum
of the operator $O$, in an $L_{S}-$periodic lattice. Thus, when
$O=m,$ one gets excitations from a magnetic system or, with $O=L$,
the spectrum of screening-masses for the gluonic field is recovered.
While in the first case the spectrum is generated by a magnetic phase-transition
that breaks the global $Z_{2}-$symmetry, in the latter case it is
due to a spontaneous $Z_{2}$-center symmetry breaking of the $SU(2)$
gauge group \cite{Svet&Yaffe,Fiore,Falcone}.

Although inherently different in nature, the aforementioned phase-transitions
constitute critical phenomena in the same universality class \cite{Stat_Field_Theory}.
Thus, one may expect that universality \cite{Svet&Yaffe} can predict
some dynamical aspects of gauge theories, such as mass-ratios \cite{Crit_Phen_Vicari}
of excitations in Eq.(\ref{Corr-Matrix}), to be shared with statistical-mechanical
systems. While evidences supporting this hypothesis are accumulating
\cite{Fiore,Falcone}, arguments entirely based on universality are
not enough to describe deconfinement in sympletic (or exceptional)
gauge theories \cite{GeneralGroups}. Thus, a broader dynamical picture
may be needed \cite{Mapping_Theorem_Frasca}.

An improved understanding of such dynamics may be obtained by comparing
the spectrum of gauge theories and exactly-solvable systems in the
same universality class. For instance, the $2d$ Ising model is well-described
by a free-fermion field theory near criticality, whose spectrum of
excitations --- for $n$ integer --- turns to be\begin{equation}
M_{n}=nM_{0},\label{Ising_Spectrum}\end{equation}
where $M_{0}$ is a mass-gap proportional to $|\tau|$: the deviation
from critical temperature \cite{IsingSpectrum}. At the same time,
the spectrum of $\lambda\phi^{4}$ theory is analytically calculated
by a gradient-expansion \cite{Frasca_Phi4_Spectrum}, which results
with mass-scales $(\Lambda_{broken})$ --- and integer $k$ --- for
the broken phase \begin{equation}
M_{k}=k\Lambda_{broken}.\label{Phi_Spectrum_Broken}\end{equation}

Therefore, one can apply Monte Carlo methods to exploit whether the
screening spectrum of the YM theory matches expectations from Eq.(\ref{Ising_Spectrum})
and Eq.(\ref{Phi_Spectrum_Broken}) near criticality, a regime where
correspondences are expected, but other non-perturbative techniques
are less effective (since the theory is dynamically trivial just in
2d; i.e., at the $T\rightarrow\infty$ limit \cite{Sorella_2d_and3d4d,Maas_2d}).

\section{Algorithms }

Since in the vicinities of phase-transitions the CSD phenomena \cite{Sokal_Wolff}
afflicts the generation of statistically independent gauge-configurations
more severely, continuous efforts toward the improvement of thermalization
algorithms are ubiquitous for Monte Carlo simulations. To ensure an
efficient thermalization of gluonic fields, we applied our modified
heat bath algorithm (MHB), which was shown to be faster \cite{MHB}
than the usual heat bath (HB) algorithm \cite{Creutz,KennedyPendleton}. 

Usually, when an $SU(2)$ lattice-gauge theory is considered \cite{MHB},
one can factorize the action Eq.(\ref{Wilson_Action}) as a sum over
many single-link actions $S_{1-link}$ as

\begin{equation}
S_{1-link}=-\frac{\beta}{2}Tr\left[U_{\mu}\left(x\right)H_{\mu}\left(x\right)\right],\label{S_1link}\end{equation}
with $U_{\mu}\left(x\right)\in SU\left(2\right);$ $H_{\mu}\left(x\right)=N_{\mu}\left(x\right)\tilde{H_{\mu}}\left(x\right)$
is a sum over staples --- i.e., it is proportional to an SU(2) matrix
--- where $\tilde{H_{\mu}}\left(x\right)\in SU\left(2\right)$ and
$N_{\mu}\left(x\right)=\sqrt{\det H_{\mu}\left(x\right)}$. 

Then, by using Eq.(\ref{S_1link}) and the invariance of the group
measure under group multiplication \cite{Creutz}, the usual HB update
is obtained\begin{equation}
U_{\mu}^{old}\left(x\right)\longrightarrow U_{\mu}^{new}\left(x\right)=V\tilde{H_{\mu}^{\dagger}}\left(x\right),\label{HB_update}\end{equation}
 where $V=v_{0}I+i\vec{\cdot v}\cdot\vec{\sigma}\in SU\left(2\right)$
is randomly generated by choosing $v_{0}$ distributed as $\sqrt{1-v_{0}\lyxmathsym{\texttwosuperior}}$$\exp\left(\beta Nv_{0}\right)dv_{0}$
and $\vec{v}$ is \textit{randomly} chosen in $\mathbb{R}^{3}$. Our
algorithm \cite{MHB} proceeds as the HB algorithm to generate the
updating matrix $V$, except for the additional step
\begin{itemize}
\item \textit{Transform the new vector-components of V as $\vec{v}\rightarrow-sgn(\vec{v}\cdot\vec{w})\vec{v}$,}
\end{itemize}
where $W=w_{0}I+i\cdot\vec{w}\cdot\vec{\sigma}=U_{\mu}^{old}\left(x\right)\tilde{H_{\mu}}\left(x\right)$,
and $sgn$ is the sign function.  

Still, MHB may be seen as a modification of the overheat-bath algorithm
(OH) \cite{OHB} that incorporates a micro-canonical move \cite{Adler}
into a heat-bath step. However, while in MHB all components of the
vector $\vec{v}$ are randomly generated (i.e., except for its sign)
in very close similarity to HB, the OH algorithm \textit{deterministically}
sets $\vec{v}=-\vec{w}$ (and renormalizes it as $\left\Vert \vec{v}\right\Vert =\sqrt{1-v_{0}^{2}}$).
Thus, OH incorporates a microcanonical move in an \textit{exact},
but \textit{maybe} non-ergodic (see for example discussions in \cite{MHB})
algorithm, which is corrected --- by construction --- in the MHB version.

\section{Spectroscopical methods}

The spectrum of excitations of a field theory on the lattice can be
directly extracted by brute-force least-squares fitting from Eq.(\ref{Corr-Matrix})
using a multiple-exponential decay function. Despite being straightforward,
this method has several drawbacks and allows for reliable results
just when high-quality statistics is available. More robust techniques
such as Bayesian fit \cite{Bayesian_Fit2} or the Evolutionary fitting
\cite{Evolutionary_Fit} would constitute alternatives to overcoming
these difficulties. However, here we use another state-of-art approach,
namely the variational method \cite{Variational_Method}, broadly
employed in hadron spectroscopy \cite{Variational_BGR}. 

On the variational method a proper set of base-operators (i.e., the
interpolators $O_{i}$) has to be chosen for building a cross-correlation
matrix \begin{equation}
C_{ij}\left(\left|r\right|\right)\equiv<O_{i}\left(r\right)O_{j}^{\dagger}\left(0\right)>-<O_{i}\left(r\right)><O_{j}^{\dagger}\left(0\right)>,\label{Cross_Corr_Matrix}\end{equation}
which may be diagonalized in a generalized eigenvalue problem (properly
normalized at some $r_{0}<L_{S}$), resulting in

\begin{equation}
C\left(r\right)\cdot\vec{v_{n}}=\lambda_{n}\left(r,r_{0}\right)\cdot C\left(r_{0}\right)\cdot\vec{v_{n}},\label{Generalized_Eval_Prob}\end{equation}
where eigenvalues behave as \begin{equation}
\lambda_{n}\left(r,r_{0}\right)\varpropto e^{-\left(r-r_{0}\right)m_{n}}\left[1+\mathcal{O}\left(e^{-\left(r-r_{0}\right)\Delta m_{n}}\right)\right].\label{Eigenvalues}\end{equation}
Generally, $\Delta m_{n}$ is the mass difference to the closest lying
state, where each interpolator $O_{i}$ (projected to defined momentum-states)
has quantum numbers in a given channel. 

Within a large enough basis of independent interpolators, each eigenvalue
of Eq.(\ref{Cross_Corr_Matrix}) will decay as the leading order of
Eq.(\ref{Eigenvalues}). Thus, the slowest decay-mode of eigenvalues
is associated with the ground state, while the fastest one gives the
highest-excitated state. This implies that simple fits of a few parameters
may be nicely done, since single-exponentials dominate the signal
in the whole range. In order to identify proper fit-ranges, it is
useful to locate stable plateaus not only for effective-masses of
$\lambda_{n}$ 

\selectlanguage{brazil}%
\begin{equation}
m_{n}^{eff}\left(r+1/2\right)=\ln\left(\frac{\lambda_{n}\left(r\right)}{\lambda_{n}\left(r+1\right)}\right),\label{EffectiveMasses}\end{equation}
\foreignlanguage{english}{but also in their associated eigenvectors
$\vec{v_{k},}$ which work as fingerprints of each state. }

\selectlanguage{english}%
We are interested in a set of interpolators that generates scalar
screening-excitations in the $0^{++}$ channel. An efficient method
for building this is to apply recursive smearing-steps over usual
wall-averaged (i.e. zero-momentum) Polyakov loop operator \cite{Fiore},
which is defined by\begin{equation}
\overline{L}\left(x\right)=\frac{1}{\left(L_{S}\right)_{y}}\sum_{n=1}^{n=\left(L_{S}\right)_{y}}L\left(x,na\right).\label{UsuaWallPolyakov}\end{equation}
It enables access to all length-scales on the lattice through the
computation of cross-correlation functions among \textit{nth-smeared}
interpolators (i.e. $P^{\left(n\right)}$), which are constructed
iteractively (for increasing smearing steps $n$) from the usual Polyakov
loop ($\overline{L}\left(x\right)$) . That operators are assembled
following the rule \foreignlanguage{brazil}{\begin{equation}
\begin{array}{c}
P^{(0)}\left(x\right)=\overline{L}\left(x\right)\\
P^{(n+1)}\left(x\right)=sign\left(u\right)\left[\left(1-\omega\right)\left|u\right|+\omega\left\langle P^{(n)}\right\rangle \right]\\
u=\frac{1}{2}\left[P^{(n)}\left(y-a\right)+P^{(n)}\left(y+a\right)\right],\end{array}\label{Smeared_Operators}\end{equation}
where $\omega\in\left(0,1\right]$} and we have taken \foreignlanguage{brazil}{$\omega=0.1$. }

\selectlanguage{brazil}%
In addition, a compromise between signal-to-noise and maximal linear
independence of correlators must be attained\foreignlanguage{english}{
\cite{Variational_Method_PruneOps}}, which is achieved by looking
for a set of interpolators that minimizes the conditioning number
($\kappa$) of the normalized correlation matrix ($\hat{C_{ij}}$)
on slice $r=r_{t}$ \foreignlanguage{english}{\begin{equation}
\hat{C_{ij}}=\frac{C_{ij}\left(r_{t}\right)}{\sqrt{C_{ii}\left(r_{t}\right)C_{jj}\left(r_{t}\right)}}.\label{NormalizedCrossMatrix}\end{equation}
Thus, noisy interpolators are eliminated by considering the signal-to-noise
ratio of the diagonal elements of Eq.(\ref{NormalizedCrossMatrix}).
Additionally, their independence is estimated remembering that $\kappa=1$
for a completely orthogonal basis (while $\kappa\rightarrow\infty$
for increasing levels of linear-dependence).}

\selectlanguage{english}%

\section{Numerical results}

In our finite-temperature simulations we have used the Wilson action
Eq.(\ref{Wilson_Action}) and asymmetric lattices $L_{S}^{2}\times L_{T}$,
where we take $L_{T}=8$ and $L_{S}=\left\{ 50,70,90\right\} $. The
critical coupling we adopted is known with high accuracy to be $\beta=12.63$
\cite{Critical_Beta_SU2_LQCD,Critical_SU2_Smekal} for this case. 

In 2+1 dimensions the gauge coupling $g_{s}^{2}$ has the dimension
of mass and sets a scale for the theory. For instance, the lattice
spacing $(a)$ can be calculated using the string-tension $(\sigma)$
--- as a numeric physical input --- on known $\beta$-string tension
relations for $SU\left(N\right)$ in three dimensions \cite{Teper}.
We have assumed $\sqrt{\sigma}=0.440$ $GeV$ as in \cite{Cucchieri_scale_3d,G2_LGT_Maas}
--- i.e. a zero-temperature $4d$ numerical value --- which gives
$a\simeq0.0503$ $fm$ for $SU(2)$. 

So, our simulations are in the asymptotic region, where discretization
effects (UV) due to coarse-lattices \cite{Strong_coupling_limit_QCD}
would be weak. We employed the MHB thermalization algorithm, which
incorporates an overrelaxation step, to generate respectively $\left\{ 5M,10M,15M\right\} $
gauge-configurations at critical temperature for lattices $L_{S}=\left\{ 50,70,90\right\} $.
The statistical independence of gauge configurations was checked by
computing the auto-correlation time $\tau_{int}$ \begin{equation}
\begin{array}{c}
\tau_{int}=\frac{1}{2}+\sum_{k=1}^{\infty}\rho_{\mathcal{O}}(k),\\
\rho_{\mathcal{O}}(k)=\frac{\left\langle \mathcal{O_{\textrm{i}}\mathcal{O_{\textrm{i+k}}}}\right\rangle -\left\langle \mathcal{O_{\textrm{i}}}\right\rangle ^{\textrm{2}}}{\left\langle \mathcal{O_{\textrm{i}}}^{2}\right\rangle -\left\langle \mathcal{O_{\textrm{i}}}\right\rangle ^{\textrm{2}}},\end{array}\label{IntegratedCorrTime}\end{equation}
for the Polyakov loop operator (i.e. $\mathcal{O}=\bar{L}$) among
successive samples. An automatic windowing procedure with $c=[4,10]$
and the Madras-Sokal formula for error estimation were employed \cite{Sokal_Wolff}.
We kept $1k$ independent gauge configurations for spectroscopical
analysis. Statistical error-bars were computed by the bootstrap method
\cite{LatticeQCD,Sokal_Wolff} with $70k$ repetitions, see Figure
(1).

The interpolators in Eq.(\ref{Smeared_Operators}) were pruned as
a function of their smearing level using the aforementioned methods.
The better suited iteraction-levels we could determine were $n=\left\{ 3,9,15,21,27\right\} ,$
with $\kappa\approx1.2$. Although up to five interpolators were used
to build cross-correlation matrices, as in Eq.(\ref{Cross_Corr_Matrix}),
no more than three mass-excitations could be recovered by our variational
approach. 

To proceed the determination of the spectrum of screening masses,
we performed least-squares global-fits to exponential decays of eigenvalues
of cross-correlation matrices assuming\begin{equation}
\lambda_{n}\left(r,r_{0}\right)=A+B_{n}\cosh\left[m_{n}\left(\frac{L_{S}}{2}-\left|r-r_{0}\right|\right)\right],\label{EigenvalFit_NoTunneling}\end{equation}
as shown in Figure (1) and Table (1). Near phase transitions, some
finite-size (and tunneling) effects were reported \cite{Fiore,Falcone}
to induce sub-leading contributions to correlators Eq.(\ref{Eigenvalues}),
which are handled by the constant $A$. Each effective-mass is obtained
from the best fit inside of plateaus, where (at least three) neighbour
masses agree within one error-bar.

\begin{figure}
\centering\includegraphics[clip,width=5.5cm]{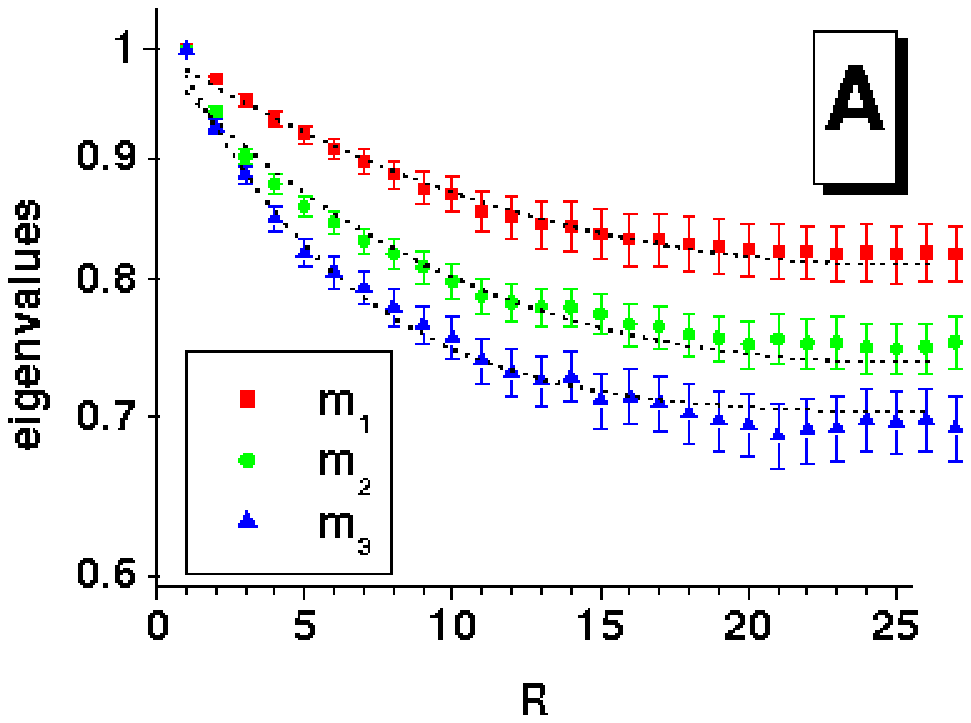}\includegraphics[clip,width=5.5cm]{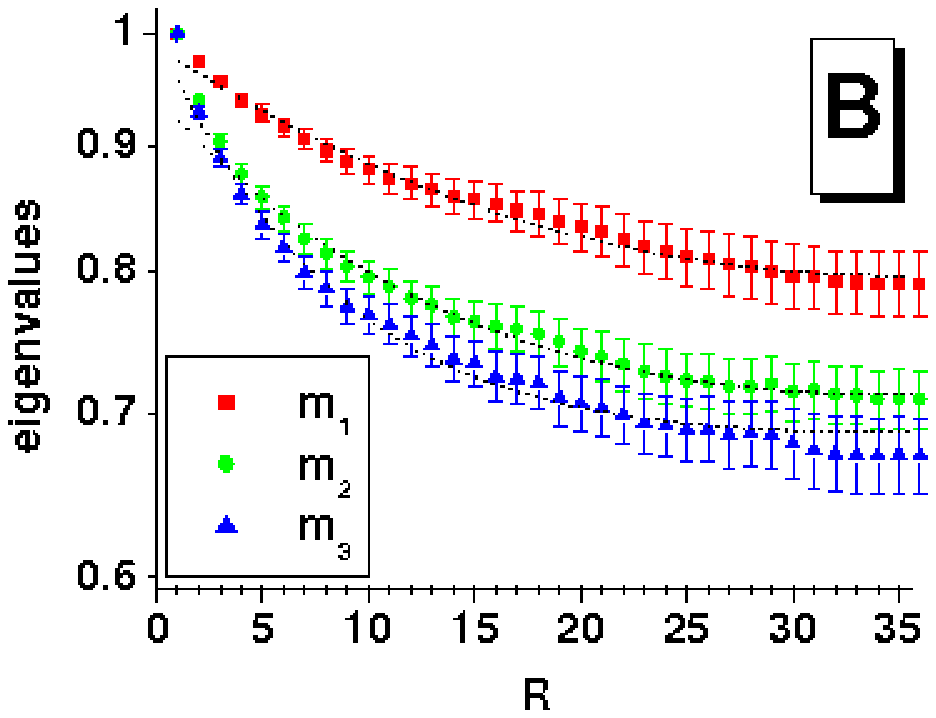}\includegraphics[clip,width=5.5cm]{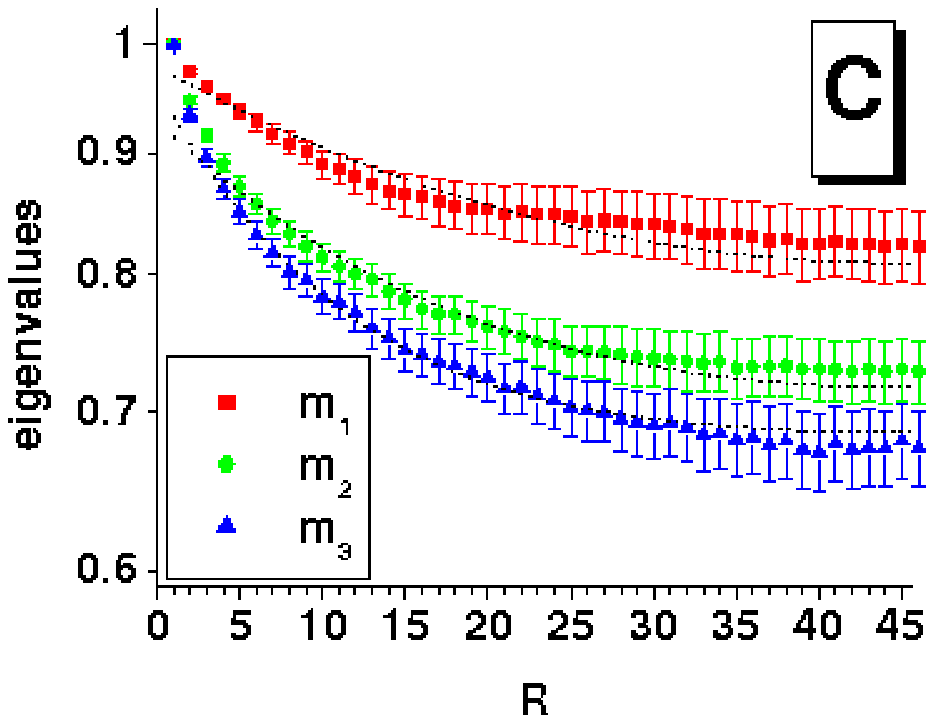}

\caption{Eigenvalues of the cross-correlation matrix for $P^{\left(n\right)}$-operators
with $n=\left\{ 3,9,15,21,27\right\} $. Panel A: $L_{S}=50$, panel
B: $L_{S}=70$, panel C: $L_{S}=90,$ }

\end{figure}

\begin{table}
\centering\begin{tabular}{|c|c|c|c|c|c|c|}
\hline 
Mass & $L_{50}$ & fit range & $L_{70}$ & fit range & $L_{90}$ & fit range\tabularnewline
\hline 
$M_{3}$ & 0.191(8) & {[}3,47{]} & 0.134(3) & {[}2,68{]} & 0.098(4) & {[}3,87{]}\tabularnewline
\hline 
$M_{2}$ & 0.107(3) & {[}2,48{]} & 0.081(2)  & {[}4,66{]} & 0.058(2) & {[}5,85{]}\tabularnewline
\hline 
$M_{1}$ & 0.065(1) & {[}2,48{]} & 0.0467(5) & {[}3,67{]} & 0.0329(5) & {[}3,87{]}\tabularnewline
\hline
\end{tabular}\caption{The screening-masses in lattice units (Mass) obtained by fitting the
eigenvalues of the cross-correlation matrix of $P^{n}$-correlators
by Eq.(\ref{EigenvalFit_NoTunneling}), with $n=\left\{ 3,9,15,21,27\right\} $
for lattices $L_{S}=\left\{ 50,70,90\right\} $. The ranges shown
correspond to best-fits with goodness $\left(\chi\lyxmathsym{\texttwosuperior}/dof\right)_{L50}=0.306,$
$\left(R\lyxmathsym{\texttwosuperior}\right)_{L50}=0.993$; $\left(\chi\lyxmathsym{\texttwosuperior}/dof\right)_{L70}=0.139,$
$\left(R\lyxmathsym{\texttwosuperior}\right)_{L70}=0.997$ and $\left(\chi\lyxmathsym{\texttwosuperior}/dof\right)_{L90}=0.248,$
$\left(R\lyxmathsym{\texttwosuperior}\right)_{L70}=0.993$. Error-bars
were computed by $70k$ bootstrap-repetitions performed over 1000
independent configurations. Integrated auto-correlation times $\left(\tau_{int-P}\right)$
associated to Polyakov loops $(i.e.\: P^{\left(0\right)})$ were used
for checking the statistical independence of samples, they were computed
by automatic windowing procedure with $c=[4,10]$. The error estimation
for each $\tau_{int-P}$ is from Madras-Sokal formula \cite{Sokal_Wolff},
resulting in $\left(\tau_{int-P}\right)_{L50}=5366(645)$, $\left(\tau_{int-P}\right)_{L70}=9526(1013)$
and $\left(\tau_{int-P}\right)_{L90}=15108\left(1402\right)$. }

\end{table}

We also performed a finite-size scaling (FSS) study using the measured
masses ($m_{n}$) for extraction of their infinite-volume limit \cite{Potts_Alves}.
A fitting ansatz $m_{n}=a_{n}/L+b_{n}$ suggests $b_{n}\equiv0$ for
all masses, while $m_{n}=a_{n}L^{-\nu_{n}}$ corroborates with $v_{n}\equiv v$,
within one standard-deviation. Thus, the following functional dependence
\begin{equation}
m_{n}=a_{n}L^{-\nu},\label{Mass_Scalling}\end{equation}
was employed for fitting simultaneously all data points, as is seen
in Figure (2).

The obtained exponent $\nu=1.16(6)$ shows that at IR limit all mass-excitations
$\left(m_{n}\right)$ scale as the inverse correlation-length $\left(\xi\right)$,
as in the critical 2d Ising-model \cite{Crit_Phen_Vicari}. The average
ratios computed from face values in Table (1), using error-propagation
formulas, are compatible with results calculated from coefficients
$a_{n}=[a_{1}=6.4(1.7),$ $a_{2}=10.7(2.9),$ $a_{3}=18.4(5.0)]$
from the best-fit (i.e. $\chi\lyxmathsym{\texttwosuperior}/dof=0.808,$
$R\lyxmathsym{\texttwosuperior}=0.993)$ with Eq.(\ref{Mass_Scalling}).
Thus, it is pointed out that ratios for screening-masses in $0^{++}$
channel of the critical finite-temperature YM theory in $\left(2+1\right)d$
are $m_{2}/m_{1}=1.69(15)$ and $m_{3}/m_{1}=2.89(35)$.

Hence, the measured spectrum can be fitted by $b_{n}\equiv0$, a behavior
associated with a tower of massless excitations is seen at thermodynamic
limit. Morever, our data agree with universality arguments Eq.(\ref{Ising_Spectrum})
and, so far, the prediction Eq.(\ref{Phi_Spectrum_Broken}) from mapping
\cite{Mapping_Theorem_Frasca} stays valid for $d\neq4.$

\begin{figure}[H]
\centering\includegraphics[clip,width=7cm]{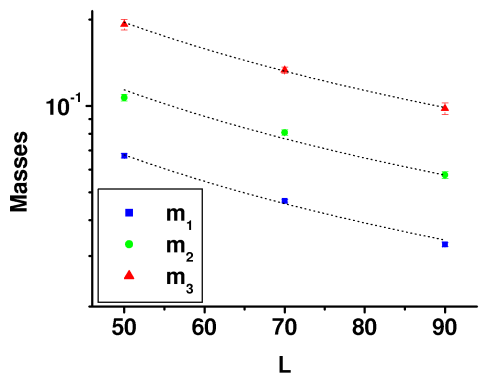}

\caption{Scaling of effective-masses as a function of lattice sides $(L)$.
Dot-line is the best global-fit from Ansatz in Eq.(\ref{Mass_Scalling}),
its goodness is $(\chi\lyxmathsym{\texttwosuperior}/dof=0.808,$ $R\lyxmathsym{\texttwosuperior}=0.993)$.
The exponent obtained $\nu=1.16(6)$ and the $\left(a_{n}\right)$
coefficients imply $0^{++}$ screening-mass ratios compatible with
$m_{2}/m_{1}=1.69(15)$ and $m_{3}/m_{1}=2.89(35)$.}

\end{figure}

\section{Concluding remarks}

We have computed the gluonic spectrum of screening masses (in $0^{++}$
channel) of quenched $3d$ Yang-Mills theory at critical temperature.
State-of-art techniques from hadron-spectroscopy, such as the variational-method
and recursively-smeared interpolators, were employed with our new
thermalization algorithm, to ensure accurate results at the critical
region. While the measured spectrum is expected to present no significant
systematic effects from UV-cutoff --- since we employed very fine
lattices --- a noticeable dependence on the spatial lattice side (i.e.
on the dimensionless quantity $L_{S}$) was observed.

Therefore, in the way to obtain the mass-excitations in the thermodynamic
limit a FSS extrapolation was performed, which unveiled a tower of
massless excitations, as expected by dynamical \cite{Mapping_Theorem_Frasca,Frasca_Phi4_Spectrum}
and universality-based arguments \cite{Svet&Yaffe,Fiore,Falcone,IsingSpectrum}
(i.e. similar to the spectrum of scalar/free-fermion field theories).
A result that strongly resembles patterns from bound-states of glueballs
analytically calculated for $SU(N)$ theories, at the $N\rightarrow\infty$
limit \cite{LargeN_Glueballs}.

Hence, traditional criteria for confinement are based on the gauge-dependent
behavior of gluon-propagators at the infrared limit \cite{Alkofer_Confinement};
connecting finite-\textcompwordmark{}temperature studies in different
gauges \cite{Cucchieri_Screening} to gauge-independent \cite{Maezawa_Screnning_Masses}
results is worthwhile. An analytical development in this direction
was introduced in \cite{Mapping_Theorem_Frasca}, where a mapping
from $SU(N)$ gauge theories to $N-$scalar fields consistently fits
$4d$ gluon-propagators at deep infrared. Moreover, it agrees with
modern findings concerning YM theories for more general gauge groups
\cite{GeneralGroups}, where a subtle dynamical balance among group
generators and the size of group center seems to lead to occasional
universal behavior.

Whether such assertions are valid for the finite-temperature $3d$
YM theory near criticality is a matter for numerical verification.
So, we have computed mass-ratios of $SU(2)$ gauge theory and compared
them to ratios from universality-related models, namely the $\lambda\phi^{4}$
and Ising-field theories. Our results match both universality and
dynamical-mapping hypothesis \cite{Svet&Yaffe,Fiore,Mapping_Theorem_Frasca,Frasca_Phi4_Spectrum}
in Eq.(\ref{Ising_Spectrum}) and Eq.(\ref{Phi_Spectrum_Broken});
which harmonizes with confinement-scenarios \cite{Sorella_2d_and3d4d,Maas_2d,EJPC_MassiveG=Confinement}
presenting (confined) massive gluonic-excitations in $d>2$.

For future research we leave the computation of screening-masses for
generalized YM theories \cite{GeneralGroups} using smeared interpolators
like Eq.(\ref{Smeared_Operators}), though built upon order parameters,
as the dressed Polyakov-loop, associated with chiral Dirac operators
\cite{Variational_BGR,Polyakov_Gattringer}. The resulting spectral
patterns arising from such investigations could be straightforwardly
related to the behavior of usual YM propagators by methods presented
in \cite{G2_LGT_Maas}, which may bring further insights into dynamical
aspects connecting universality to chiral and deconfinement phase-transitions
\cite{QCD_Universality_O4}.
\begin{ack}
The author thanks Tereza Mendes and Attilio Cucchieri for drawing
his attention to this problem and, together with Marco Frasca, for
enlightening discussions. Financial support was provided during successive
stages by FAPESP and CAPES (Brazil).\end{ack}

\end{document}